\nonstopmode

\documentclass[pre,twocolumn]{revtex4}
\usepackage{graphicx}
\usepackage{natbib}
\usepackage{epstopdf}
\usepackage[hidelinks, colorlinks = true, urlcolor=blue, linkcolor=blue, citecolor= blue]{hyperref}
\usepackage{color}
\usepackage{amsmath}
\usepackage{lettrine}
\usepackage{lipsum}
\begin{document}

\title{Durable Bistable Auxetics Made of Rigid Solids}

\author{Xiao Shang$^1$, Lu Liu$^1$, Ahmad Rafsanjani$^2$ and Damiano Pasini$^{1}$}
\email[Correspondence to ]{damiano.pasini@mcgill.ca}

\affiliation{$^1$Department of Mechanical Engineering, McGill University, Montreal, Quebec, H3A 0C3, Canada}
\affiliation{$^2$John A. Paulson School of Engineering and Applied Sciences, Harvard University, Cambridge, MA 02138, USA}

\date{\today}

\hyphenpenalty=5000

\begin{abstract}

Bistable Auxetic Metamaterials (BAMs) are a class of monolithic perforated periodic structures with negative Poisson’s ratio. Under tension, a BAM can expand and reach a second state of equilibrium through a globally large shape transformation that is ensured by the flexibility of its elastomeric base material. However, if made from a rigid polymer, or metal, BAM ceases to function due to the inevitable rupture of its ligaments. The goal of this work is to extend the unique functionality of the original kirigami architecture of BAM to a rigid solid base material. We use experiments and numerical simulations to assess performance, bistability and durability of rigid BAMs at 10,000 cycles. Geometric maps are presented to elucidate the role of the main descriptors of BAM architecture. The proposed design enables the realization of BAM from a large palette of materials, including elastic-perfectly plastic materials and potentially brittle materials.
\end{abstract}

\maketitle

%============== Introduction

\section{INTRODUCTION}
Metamaterials are designer matter with unusual properties that are governed by their underlying architecture rather than their chemical composition. Firstly introduced in the field of electromagnetic materials~\cite{Pendry1999}, the concept quickly extended to mechanical systems~\cite{Lee2012}. Since then, a great interest in the mechanics of materials arena has emerged due to the exotic and tunable properties that mechanical metamaterials can offer, such as negative modulus and mass density, vanishing shear modulus, and programmable properties among many others~\cite{Ding2007, Liu2011, Kadic2012, Milton1992, Rafsanjani2015, Haghpanah2016}. 
The Poisson’s ratio, ν, the ratio between the transverse and longitudinal strain in the direction of an applied stretch, is a property of interest here; for isotropic 3D materials $\nu$ can vary between -1 and 0.5~\cite{Greaves2011}, whereas for 2D materials $-1<\nu<1$; for auxetics it is negative. Compared to solids with positive ν, auxetic materials offer enhanced shear moduli, indentation resistance, and fracture toughness~\cite{Evans1991,Lakes1987}, thus providing mechanical gains in a large range of applications, such as medical stents and skin grafts~\cite{Ali2014,Gatt2015}, porous dampers and acoustic foams~\cite{Scarpa2004}, as well as gas turbines~\cite{Taylor2014,Javid2017}. Several efforts have been recently devoted towards the creation of auxetic sheets obtained by perforating monolithic sheets with uniformly~\cite{Grima2010}, randomly~\cite{Grima2016}, as well as hierarchically~\cite{Gatt2015} distributed incision patterns. In three-dimensions auxetic materials have been also obtained from re-entrant origami-based Tachi-Miura polyhedrons~\cite{Yasuda2015} and origami tubes~\cite{Filipov2015}. Other works on auxetic materials exploit elastic instability to modulate pattern switch and trigger distinct shape transformations. These include square planar lattices with buckled circular holes under compression~\cite{Bertoldi2010}; metallic metamaterials designed to exploit instability in regular planar lattices~\cite{Ghaedizadeh2016}; three-dimensional auxetics made up of sinusoidal buckling elements~\cite{Findeisen2017}; and three-dimensional soft auxetics~\cite{Babaee2013}. Whereas all these efforts demonstrate that elastic instability locally distributed in a monolithic material can bring about negative Poisson’s ratio, all of them return to their original shape once the elastic strain is removed. Multistable materials, on the other hand, can assume besides their initial position, additional configurations of equilibrium~\cite{Rafsanjani2015, Haghpanah2016, Prasad2005, Restrepo2015}, but seldom do they exhibit auxeticity. So far in the literature, only few works combine auxeticity and multistability in monolithic materials~\cite{Hewage2016, Rafsanjani2016}, with examples created mainly from origami patterns~\cite{Yasuda2015, Silverberg2015}.

%============== Fig 1
\begin{figure}[b]
\centering
\includegraphics [width=\columnwidth]{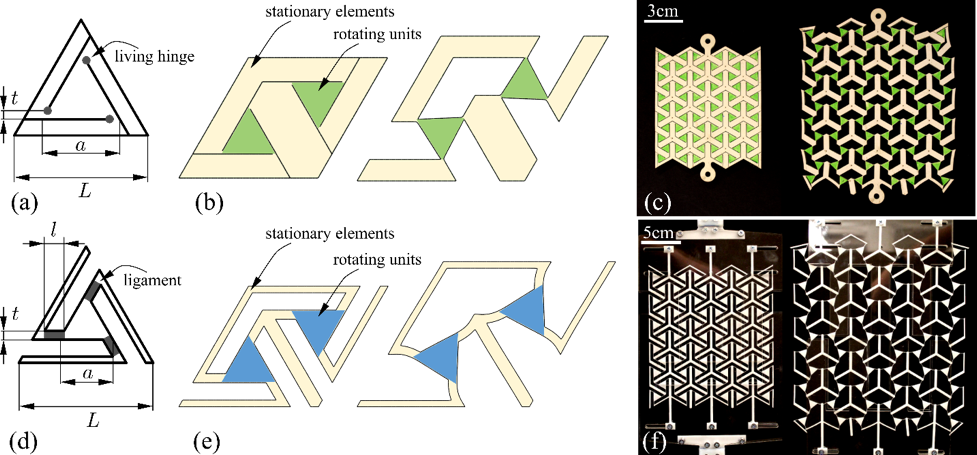}
 \caption{Original BAM Architecture: (a) building block geometry, (b) a unit cell in its closed and open state (c) a natural latex rubber sample with $6\times6$ modules in its undeformed and fully open configuration~\cite{Rafsanjani2016}. Cellular BAM made of rigid solid: (d) building block geometry, (e) a unit cell in its closed and open state (f), a cellular BAM sample with $6\times6$ modules fabricated from Delrin$^\text{\textregistered}$ Acetal Resin in its undeformed and fully open configuration. The sample is sandwiched between two Plexiglas plates to prevent out of plane deformation during tensile testing.}
\label{Fig1}
\end{figure}

Bistable Auxetic Materials, in short BAMs, represent a class of monolithic perforated metamaterials with combined negative ν and pattern switch functionality, making them capable of expansion and contraction between two states of equilibrium. BAM can meet desired expandability, stiffness and bistability, at varying levels; they can do so by exploiting the functionality of the building block which is engineered from a set of slits with given rotational symmetry and well-defined geometric descriptors~\cite{Rafsanjani2016}. The constituent solid BAM has been so far made of is an elastomer, whose constitutive relation enable shape transformations that are reversible and repeatable with high compliance and quite low strength. Fig.~\ref{Fig1}(a) shows one example. Here the building block has point-wise living hinges (gray points) that connect rotating units [green in Fig.~\ref{Fig1}(b)] and stationary elements [pale yellow in Fig.~\ref{Fig1}(b)] working mainly as rigid bodies. Under an applied stretch, the rotating units rotate with respect to the living hinges, a mechanism that expands BAM to its opening state [Fig.~\ref{Fig1}(b)]. This pattern switch is repeatable and it is ensured by the elastomer properties. With a rigid solid, such as a metal, however, BAM can no longer function because its hinges inevitably fracture at the first attempt to opening it. Realizing BAM made of rigid solids, on the other hand, can be an asset in several applications, because rigid solids can offer high melting and softening temperature, along with high strength and stiffness, characteristics that are sought after in aerospace applications, such as turbo engines and solar panels. 
The goal of this work is to extend the range of materials BAM can be made of to rigid materials without losing its hallmark function, i.e. reversible and repeatable switch of opening and closure. To give a context to this characteristic, we use here the term durability, which we define as the resistance to sustain more than 10,000 cycles of opening and closure. This function is considered preserved if BAM either responds elastically or experiences a moderate level of yield in very confined regions. 10,000 cycles is representatively chosen as our target value because such a number of cycles is considered sufficiently high for certain applications, such as mechanical switches~\cite{Jensen1999} and deployable aerospace components~\cite{Hanaor2001, Zhao2009}. In Sec.~II of this paper, the perforated pattern of the original BAM is extended by introducing slender ligaments in well-defined gaps bridging rigid bodies that can either rotate or simply translate. The results are then discussed in the following sections, where we present parametric charts defining a generalized geometry space that greatly expands the range of raw material selection for BAM.

\section{METHODOLOGY}
\subsection{BAM architecture: a compliant extension of the original pattern} 
Fig.~\ref{Fig1}(a) to (c) illustrate one, among others, original BAM obtained by incising a planar sheet of natural latex rubber. The building block, shown in its closed and open configuration along with the monolitic perforated assembly of $6\times6$ units, comprises three mutually intersecting lines nesting an equilateral triangle within a triangular unit. Other unit types are also possible on square grids. The pattern in Fig.~\ref{Fig1}(a) can work smoothly with an elastomeric material only. The original BAM can be considered as a kirigami architecture with living hinges localized at physical points and without any voids between building blocks in the undeformed state. To realize a BAM with a rigid solid, we present here a cellular alteration of the original pattern [Fig.~\ref{Fig1}(d), (e) and (f)] that can sustain repeated cycles of opening and closure. The difference between the two, i.e. original and modified BAM, is that the former embeds point-wise hinges undergoing large strain, whereas the latter replaces them with slender compliant ligaments that define a distinct pattern of voids in the initial configuration. As a representative rigid solid, we chose here Delrin$^\text{\textregistered}$ Acetal Resin for its constitutive law, which is linear elastic-perfectly plastic resembling that of a metal, and for its easiness to perforate with a laser cutter. 
Fig.~\ref{Fig1}(d) and~(e) show the new geometric parameterization for the cellular BAM ligaments, all sharing length ($l$) and thickness ($t$). The generalized pattern here proposed is scalable in that the rotating unit length ($a$), the ligament length ($l$) and the ligament thickness ($t$) are all normalized by the edge length ($L$). For all samples investigated in this study, the normalized length of the rotating unit is representatively chosen as $a/L = 0.4$, and for this particular value we study the role of the normalized ligament length ($l/L$) and ligament thickness ($t/L$), both controlling bistability levels and failure conditions. The unit cell has edge length $L$ = 36mm, $l/L = 0.14$ and $t/L = 0.113$. The material properties of Delrin$^\text{\textregistered}$ Acetal Resin are obtained following the standard ISO 527-1:2012 applied to the testing of dog-bone samples with a thickness of 1.6mm, from which the following values were obtained: $E = 3.85$ GPa for the Young’s modulus, $\sigma_y = 41.18$ MPa for the yield strength, $\varepsilon_y = 0.012$ for the yield strain and $\sigma_{ult} = 80.16$ MPa for the ultimate strength. Fig.~\ref{Fig1}(f) shows a sample with 66 units, along with the special fixture that was designed for BAM to constrain its out-of-plane deformation during expansion while allowing lateral expansion caused by the negative Poisson’s ratio of the metamaterial. $6\times6$ units were chosen because this number of units was proved capable of providing BAM with a smooth and reversible opening and closing.

\subsection{Cellular BAM ligaments}
\begin{figure}[t]
\centering
\includegraphics [width=\columnwidth]{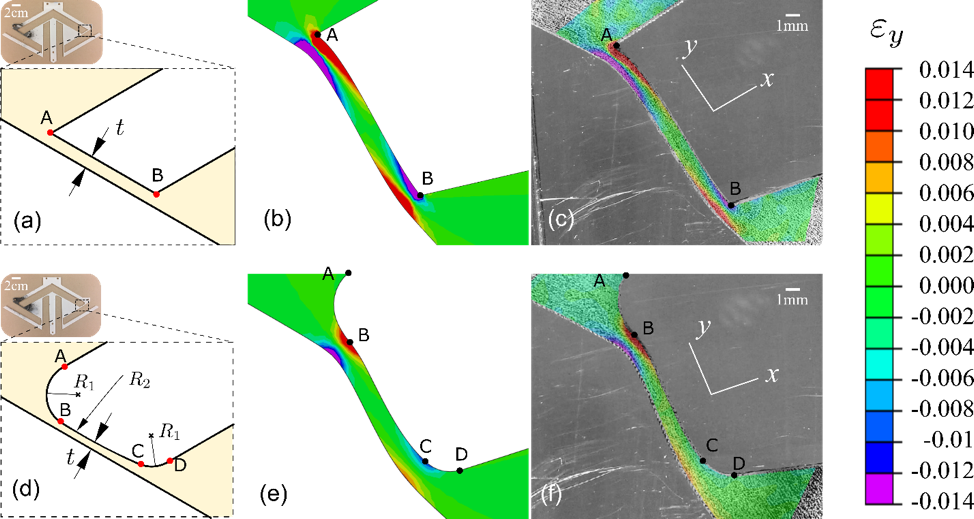}
\caption{Design of compliant ligaments for rigid BAM structures. (a) G$^0$ straight ligament with sharp edges and (d) Arc ligament with G$^1$ optimized fillets. Blending points are highlighted in red, and dark portions of ligaments in insets (a) and (d) are speckled regions used for DIC. (b, e) Finite element results for strain distribution in the direction of the ligament longitudinal axis y are compared with (c, f) experimental strains obtained by Digital Image Correlation (DIC) for G$^0$ and G$^1$ designs, respectively.}
\label{Fig2}
\end{figure}

To realize a functional BAM with a periodic pattern of voids in a rigid planar solid, we first focus on the regions subjected to localized peaks of stress, i.e. the slender ligaments [in gray in Fig.~\ref{Fig1}(d)]. For these ligaments, we study two compliant geometries, each defined by the slender beam profile and the fillets attaching to the rotating units. The first has a straight profile with sharp corners that are G$^0$ continuous [Fig.~\ref{Fig2}(a)], where G represents {\it geometry} and the superscript indicates the degree of continuity of the curve. The second ligament type has a tailored geometry with curve profile and arc fillets satisfying G$^1$ continuity at the blending points [Fig.~\ref{Fig2}(d)], whereby G$^1$ refers to the continuity of the first derivative along the inner boundary. In a straight ligament with sharp edges [Fig.~\ref{Fig2}(a)], G$^0$ guarantees continuity only between the geometric primitives at the blending points A and B. On the other hand, the curved ligament with G$^1$ optimized fillets [Fig.~\ref{Fig2}(d)] features continuous tangent at the blending points (B and C) between the arc ligament, BC, and the adjacent arcs, AB and CD. Alternating the straight ligament AB in Fig.~\ref{Fig2}(a) with the shallow arc BC with radius R2 in [Fig.~\ref{Fig2}(c)] is beneficial for an increase of BAM durability. In particular to further do so, we tailor R2 and R1 so as to minimize the maximum von Mises stress under a set of inequality constraints applied to the ligament geometry. In this process, we note that since the inner profile of the ligament (BC) is a quadratic polynomial, the thickness is not uniform, and t here refers to the average of the varying thickness between B and C. Further improvement of BAM durability can be potentially achieved by engineering G$^2$ optimized fillets~\cite{Masoumi2013}, a path that is not pursued here.

%============== Fig 2

A set of numerical analyses and uniaxial tensile tests were undertaken on samples with $L = 144$mm, $l/L = 0.14$, $t/L = 0.013$, with both G$^0$ and G$^1$ ligaments to monitor the strain in the direction of the ligament ($\varepsilon_y$). The numerical results are shown in Fig.~\ref{Fig2}(b) and (e) while Fig.~\ref{Fig2}(c) and (f) show results obtained via digital image correlation (DIC). Red and purple represent area where the strain exceeds the yield strain of the solid material, i.e.~$\varepsilon_y= 0.012$. The strain in the ligament with G$^1$ optimized fillets can redistribute more evenly with marginal areas of yield (about 8\% of the ligament surface) at the blending points [Fig.~\ref{Fig2}(d)], as opposed to the larger areas (about 40\% of the ligament surface) appearing at the sharp edges of the G$^0$ ligament. We note that these results are representative of ligaments with a given set of geometric parameters here selected to show the relative gain between the two ligament types; for samples with other geometric parameters larger areas can appear with strain above the yield but the relative difference in strain does not vary significantly.
By comparing Fig.~\ref{Fig2}(b), (e) with Fig.~\ref{Fig2}(c), (f), we find that the strain obtained from the numerical analysis parallel well the strain experimentally measured; this is valid for both the strain distribution as well as the numerical value that the strain assumes in each point. The strain distribution on the sample edges in Fig.~\ref{Fig2}(c) and~(f) is blank because DIC was unable to capture it. The good match between numerical results and experimental data further corroborate the validity of the numerical models here presented, the results of which are discussed in the following section.

%============== Fig 3
\begin{figure}[t]
\centering
\includegraphics [width=\columnwidth]{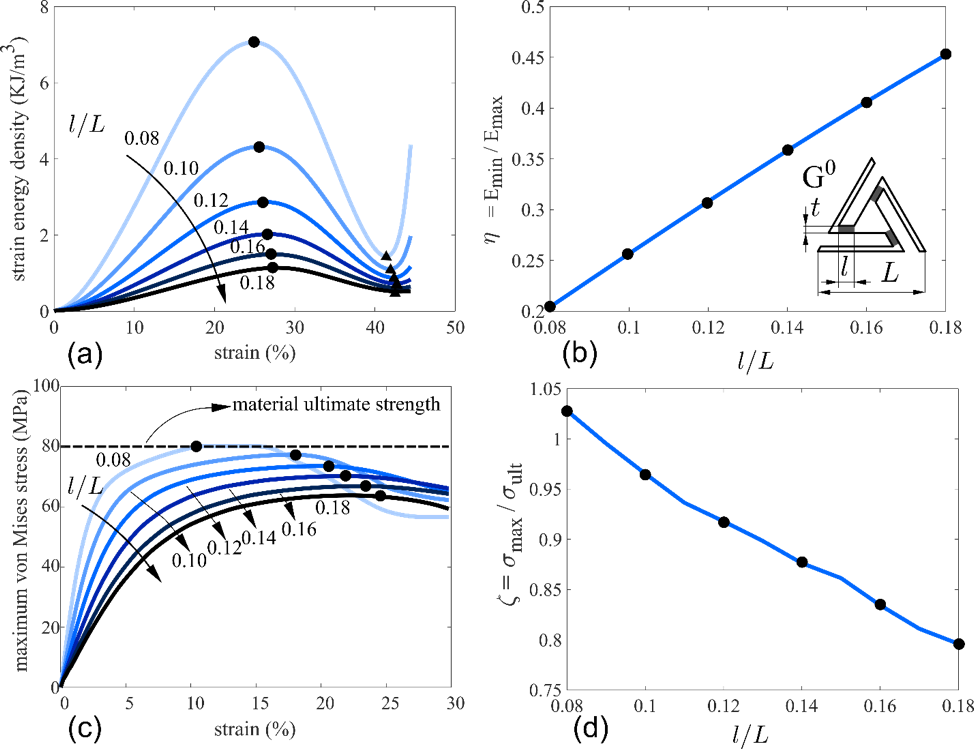}
\caption{Bistability curves and failure plots for cellular BAM with straight ligaments and sharp edges (G$^0$). (a) strain energy density versus strain for unit cells with given values of $l/L$ under periodic boundary conditions, (b) $\eta$ versus $l/L$ showing the change in bistability. (c) maximum von Mises stress versus strain for unit cells with given values of $l/L$ under periodic boundary conditions, (b) $\xi$ versus $l/L$ showing the change in failure.}
\label{Fig3}
\end{figure}

\subsection{Bistability and failure analysis for rigid BAM} 
To study bistability and failure conditions, we performed a set of nonlinear numerical analyses in ABAQUS 6.11 (Dassault Syst\`emes, France) Standard Implicit Dynamics solver on models pulled uniaxially until full expansion. Periodic boundary conditions were applied to the unit cells (Fig.~\ref{Fig1}(e)) and modified quadratic plane stress elements (CPS6M) were used to discretize the models. A contact law with a hard contact is adopted in the numeric models to represent a normal direction behavior and a frictionless tangential direction behavior. In addition, no penetration is allowed at each constraint location. ABAQUS scripts are developed to generate parametric models that are used to study the role of $l/L$ and $t/L$ on bistability and failure.

To examine how ligament length $l/L$ affects structural bistability, the strain energy density is obtained for different $l/L$. In particular, we calculate the level of bistability, $\eta = E_{min}/E_{max}$, the ratio of the local minimum strain energy pertinent to the deformed stable state of BAM [$E_{min}$, black triangular bullets in Fig.~\ref{Fig3}(a)] to its strain energy barrier [$E_{max}$, black circular bullets in Fig.~\ref{Fig3}(a)]. We note that for bistable configurations $\eta \in [0,1)$  where smaller values infer to stronger bistability.
With respect to failure, the maximum von Mises stress ($\sigma_{max}$) is calculated on the ligaments throughout BAM expansion to understand how ligament length $l/L$ controls failure. The detailed results discussed in Sec.~III are visualized in contour plots [Fig.~\ref{Fig3}(c)] of $\xi = \sigma_{max}/ \sigma_{ult}$, the ratio between $\sigma_{max}$ (black circular bullets) and the base material ultimate strength ($\sigma_{ult}$, the gray horizontal line). The ratio $\xi$ is used to assess the failure condition of rigid BAM, i.e., only for $\xi$ below 1, $\sigma_{max}$ is lower than $\sigma_{ult}$ and the rigid BAM can fully expand without structural failure.

%============== Fig 4
\begin{figure}[t]
\centering
\includegraphics [width=\columnwidth]{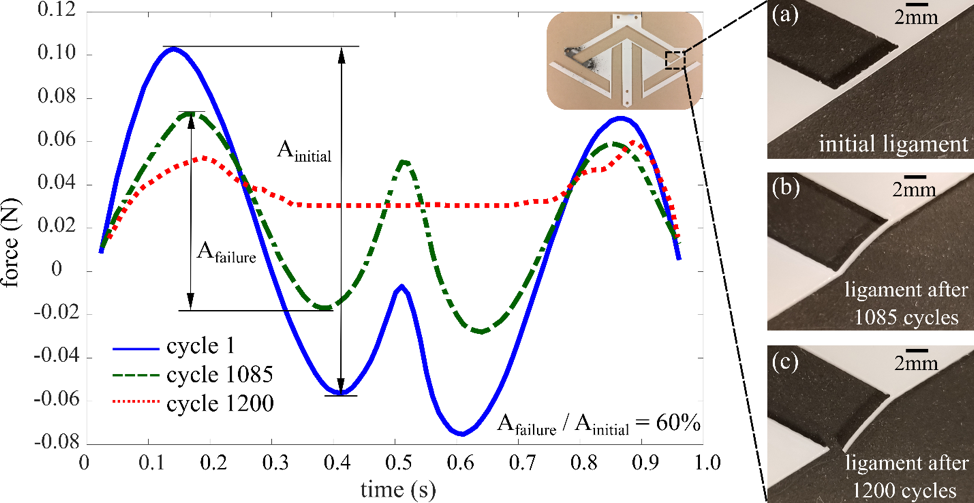}
\caption{Force-time curves for a representative unit cell at the 1st (solid line), 1085th (dashed line) and 1200th (dotted line) cycle. (a) initial ligament, (b) ligament geometry after 1085 cycles, (c) broken ligament at 1200 cycles.}
\label{Fig4}
\end{figure}

\subsection{Durability test}
The durability of the rigid BAM structures is assessed as the number of cycles a sample can resist prior to structural failure. We remark that because of the geometric peculiarity of this architected material, we did not perform a standard fatigue test. Most of previous studies on cyclic testing of architected materials were conducted under uniaxial stress or force control, whereas here BAM expansion is controlled by large displacement. In addition, the ASTM standard procedure for strain-controlled cyclic testing (ASTM E606/E606M) requires specimens with a uniform gage section or an hourglass shape. For rigid BAM, such a requirement cannot be satisfied. Hence here, we have opted to test BAM under fully reversed uniaxial displacement to assess the number of cycle to failure. We then set 10,000 loading cycles as the minimum level of durability for BAM.  
BAM unit cell samples were tested under large displacement control using a Bose ElectroForce$^\text{\textregistered}$ 3510 tester (Bose Corporation, United States). The samples were cut via a CM1490 CO$_2$ laser cutter (SignCut Inc., Canada) from 1.6mm thick Delrin$^\text{\textregistered}$ Acetal Resin sheets, with 60\% power (100W full power) and 30mm/s speed. A saw-tooth displacement curve was used at a frequency of 1~Hz with an amplitude of 30mm to ensure that all the samples could be fully expanded. We note that the frequency used for the durability test influences the number of loading cycles a rigid BAM can resist prior to failure. A complimentary study investigating this issue on representative sample with $t/L = 0.09$ and $l/L = 0.00625$ has shown that despite the difference in frequency, all numbers of cycles prior to failure lie in a range that corresponds to the results presented in this paper for 1~Hz frequency. Hence, 1~Hz is deemed a frequency sufficiently appropriate to assess rigid BAM durability in this work. Force-time curves were recorded during each test, and the specimen failure was defined at the cycle when the peak-to-peak force amplitude drops 60\% from that of the initial cycle. The number of cycles to failure ($n$) was then recorded to evaluate BAM durability. Fig.~\ref{Fig4} depicts how the durability level is assessed. All testing samples are realized with an edge length $L=72$mm, and three force-time curves are shown for one representative unit cell with $t/L=0.005$, $l/L=0.09$, respectively. 
The ligaments of the unit cell failed after 1085 cycles, as seen by comparing the insets Fig.~\ref{Fig4}(a), (b) and (c). Each curve represents the variation of the reaction force in the uniaxial loading direction in a given cycle: the solid curve corresponds to the first loading cycle; the dashed curve corresponds to the 1085th cycle; the dot-dashed curve to the 1200th cycle. Following the first cycle, localized stress confined in small regions [see Fig.~\ref{Fig2}(c)] appears in the ligaments where the von Mises stress $\sigma$ can be larger than the yield strength $\sigma_y$ of the solid material. Despite the accumulation of plastic deformation in narrowed regions, BAM can still preserve its ability to function and sustain repeated cycles of opening and closure until the 1085th cycle [Fig.~\ref{Fig4}(b)], a condition at which the peak-to-peak amplitude, Ainitial, drops below 60\% of its initial value. The failure peak-to-peak amplitude, Afailure, is defined as the value after which fast amplification of local plasticity in the ligament occurs with subsequent ligament fracture [Fig.~\ref{Fig4}(c)]. 

%============== Fig 5
\begin{figure}[b]
\centering
\includegraphics [width=\columnwidth]{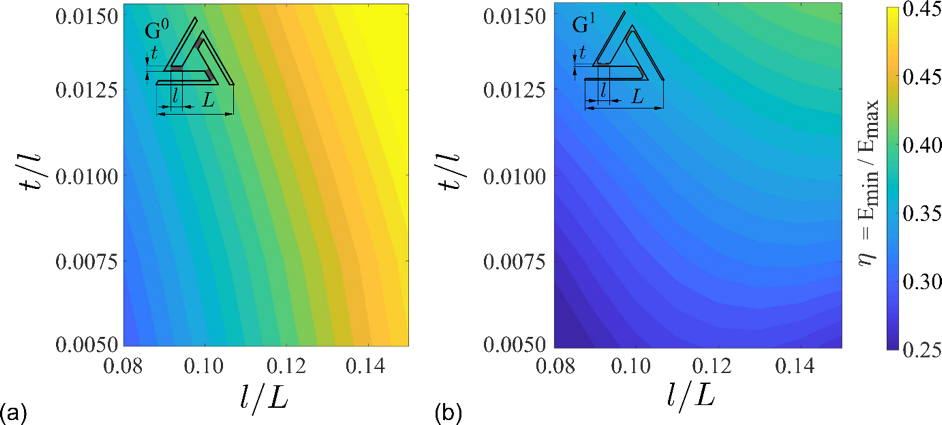}
\caption{Contour plots obtained by numerical analysis showing how bistability is controlled by $l/L$ and $t/L$ for designs with (a) G$^0$ ligament and (b) G$^1$ optimized ligament.}
\label{Fig5}
\end{figure}

\section{RESULTS AND DISCUSSIONS}
\subsection{Bistability and failure analysis}
We first focused on BAM with straight ligaments and sharp edges (G$^0$), and examined how ligament geometry influences bistability and failure. In particular, a set of numerical analyses were performed to study the impact of varying $l/L$ for given $t/L = 0.005$. The selected range of $l/L$ covers very short ($l/L = 0.08$) to long ligaments ($l/L = 0.18$). The results are plotted in Fig.~\ref{Fig3}for 6 models. We did not explore ligament length $l/L$ of 0, i.e. point-wise hinge of the original BAM, because BAM would fail instantly upon the first attempt to opening it. Numerical analysis was performed to calculate the strain energy density [Fig.~\ref{Fig3}(a)] and the maximum von Mises stress on the ligaments [Fig.~\ref{Fig3}(c)] throughout the expansion, as explained in the method section. The level of bistability, $\eta$, and failure condition, $\xi$, were calculated for each model. Shown in Fig.~\ref{Fig3}(b) and (d), the results reveal that η increases with the length of the ligament as opposed to $\xi$. An increase in ligament length results in a reduction of the amount of deformation due to ligament buckling, further leading to a reduction of $E_{max}$, the maximum strain energy. As the difference between $E_{max}$ and the minimum strain energy, $E_{min}$, reduces, the cellular BAM features a decreased level of bistability. The reduction in the amount of deformation also leads to a reduction in $\sigma_{max}$, and when $\sigma_{max}$ is below $\sigma_{ult}$, i.e. $\xi<1$, no structural failure occurs during BAM expansion. This shows that incorporating slender and compliant ligaments ensures BAM opening without incurring in structural failure.

%============== Fig 6
\begin{figure}[t]
\centering
\includegraphics [width=\columnwidth]{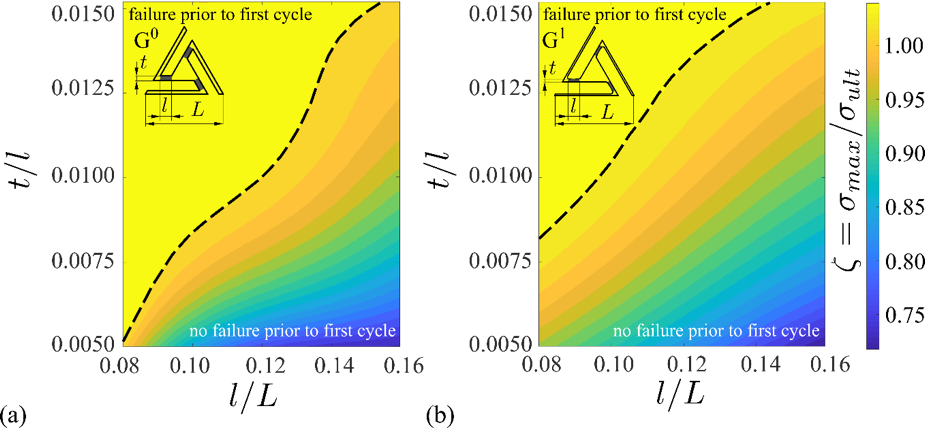}
\caption{Contour plots of failure with respect to $l/L$ and $t/L$ for (a) G$^0$ ligament and (b) G$^1$ optimized ligament.}
\label{Fig6}
\end{figure}

We also performed a set of parametric studies to investigate the sensitivity of rigid BAM bistability over the design space. Models with a set of normalized ligament length ($l/L=0.08$ to 0.14) and normalized ligament thickness ($t/L=0.005$ to 0.015) were numerically analyzed to calculate the ratio $\eta$ for each set of parameters. Fig.~\ref{Fig5}(a) and (b) show the contour plots of $\eta$ for rigid BAM with G$^0$ ligaments and G$^1$ optimized ligaments. The results indicate that an increase of $l/L$ or (and) $t/L$ corresponds to an improved level of bistability, as shown by the color gradient from dark to light. The plots also show that with a given combination of ligament length and thickness, η for a cellular pattern is smaller with G$^1$ optimized ligaments than with G$^0$ ligaments. This suggests that by modifying the ligament profile from G$^0$ to G$^1$, brings about an improved bistability. In addition, despite the presence of bistability with the parameters studied herein, we can expect that an increase of ligament length and (or) thickness would cause a transition response from bistable to monostable.
The stress peak landscape was investigated by calculating the ratio $\xi$ for models with normalized ligament length $l/L = 0.08$ to 0.14 and normalized ligament thickness $t/L = 0.005$ to 0.015. The results are illustrated in Fig.~\ref{Fig6}. The contour plots in Fig.~\ref{Fig6}(a) refer to the failure stress for BAM with G$^0$ ligaments and those in Fig.~\ref{Fig6}(b) for BAM with G$^1$ optimized ligaments. In the figures, transition curves (black dash) to failure are illustrated, i.e. curves where $\xi = 1$, below which the peak von Mises stress in the ligament allows BAM to fully expand without structural failure. By comparing the extent of the domains in Fig.~\ref{Fig6}(a) and~(b), we can observe that the use of G$^1$ optimized ligaments can reduce the peak stress in BAM. Thus, failure prior to the first cycle can be postponed by integrating G$^1$ optimized ligaments in BAM.

%============== Fig 7
\begin{figure*}[t]
\centering
\includegraphics [width=0.85\textwidth]{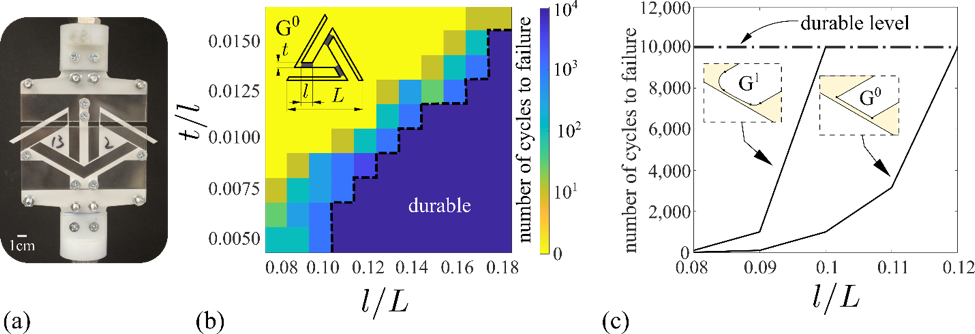}
\caption{Durability test (a) fixture designed for durability test providing the necessary boundary conditions to the unit cell sample, (b) contour plot experimentally obtained with respect to $l/L$ and $t/L$ for G$^0$ ligaments, (c) number of cycles tested to failure against l/L comparing durability level for cellular BAM samples with G$^0$ ligaments and G$^1$ optimized ligaments with a fixed $t/L = 0.0075$.}
\label{Fig7}
\end{figure*}

\subsection{BAM Durability}
A set of complementary experiments was conducted to assess the durability level of rigid BAM. A special fixture was designed to provide the necessary boundary condition for BAM expansion. In particular, the neighboring cells of the unit cell were constrained by fixing the stationary element at its two sides and its bottom, a condition that allows applying a pull at the top of the stationary element [Fig.~\ref{Fig7}(a)]. Durability tests consisted of 110 unit cell samples tested with G$^0$ ligaments throughout the design space investigated in the numerical analysis of the previous section A. For each sample, 5 specimens were tested. The contour plot in Fig.~\ref{Fig7}(b) shows the number of cycles the samples could resist prior to failure over the design space. The light-yellow area represents failed BAM, i.e. BAM ligaments broken at the first cycle, whereas the purple area represents durable BAM, i.e. BAM resists at least 10,000 cycles. Between those is a transition area with color gradient indicating levels of durability.
The durability improvement achieved by introducing G$^1$ optimized ligament was also studied experimentally. Here the test was performed for given thickness and varying length of the ligament. Fig.~\ref{Fig7}(c) shows the comparison of number of cycles to failure between BAM with G$^0$ ligament and G$^1$ optimized ligament for given $t/L = 0.0075$. Results for samples with $l/L$ (from 0.08 to 0.12) covering failure designs to durable designs with both G$^0$ ligaments and G$^1$ optimized ligaments are plotted [Fig.~\ref{Fig7}(c)]. The area above the dash-dot staircase in Fig.~\ref{Fig7}(c) is the durable area, where all BAM with longer $l/L$ can resist at least 10,000 cycles. The plot shows that with an increase of $l/L$, G$^1$ optimized ligaments allow BAM to enter the durable region much earlier than BAM with G$^0$ ligaments. In general for given $l/L$, the former resists at least one order of magnitude more cycles than the latter, a result that shows the benefit of tailoring the ligament geometry.

\subsection{Geometric maps for bistability, failure and durability}
\begin{figure}[b]
\centering
\includegraphics [width=\columnwidth]{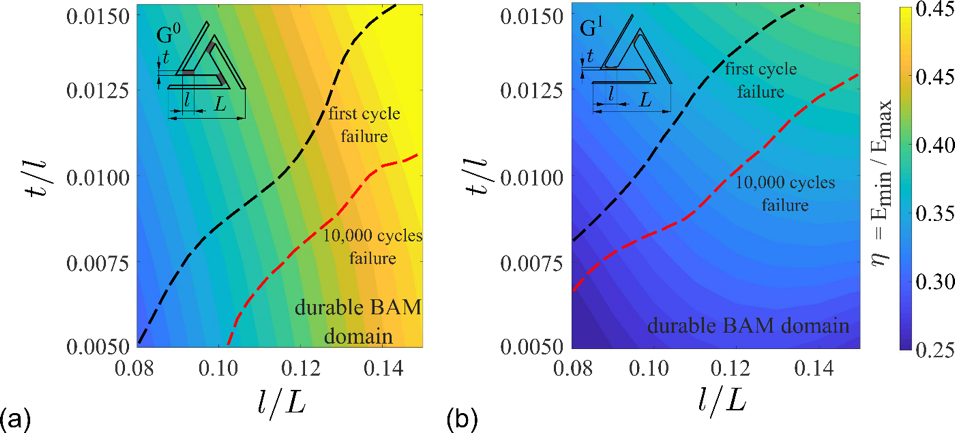}
\caption{Geometric maps illustrating how bistability, failure and durability are controlled by $t/L$ and $l/L$ for cellular patterns with (a) G$^0$ and (b) G$^1$ optimized ligaments.}
\label{Fig8}
\end{figure}

In this section, we combine the results obtained numerically and experimentally to generate maps (Fig.~\ref{Fig8}) that guide the engineering of rigid BAMs with both G$^0$ ligaments and G$^1$ optimized ligaments. Over the background spectrum describing bistability levels are superimposed two boundary curves. One in black represents ligament parameters, $t/L$ and $l/L$, for which BAM fails at the first opening; the other in red indicates BAM geometry resisting up to 10,000 cycles of opening and closure. Between them are cellular BAM which can operate for a number of cycles below the durability level set in this work. Within the design space studied in this paper, BAM geometries with best durability and the highest level of bistability are located at the bottom left corner of the durable domain. The more slender (thinner and/or longer, within the area of bistability) the ligament is, the more durable is rigid BAM.
%============== Fig 8
We remark that the maps here presented are for the nominal geometry of BAM. Manufacturing constraints might impact the selection of the best geometric parameters to perforate a rigid BAM that is functional. In addition, the geometric charts provided here are specific to Delrin$^\text{\textregistered}$ Acetal Resin, selected as representative material. The approach followed in this work however can be used to generate charts to realize compliant BAM from other elasto-plastic materials, and might be also be used for ceramic, glass and other brittle materials. 
Finally, we remark the difference between the original BAM and the concept presented in this work. The rigid BAM design proposed here is an extension of our previous work introducing BAM~\cite{Rafsanjani2016}. Both designs are fabricated from a monolithic planar sheet but there are clear distinctions between them. The original BAM is a kirigami with no gaps within building blocks, whereas the rigid BAM presented here features a cellular structure in its undeformed configuration. In addition, the original BAM made of natural rubber displays living hinges where rotation is localized at physical points, as opposed to the rigid BAM which functions via thin buckled ligaments with optimized shape profile. (For a visualization of the difference between the two BAMs, the reader is referred to the video in the supplementary material~\cite{Movie}). The concept of rigid BAM is thus a generalization of the original concept, as a reduction of the aspect ratio of the thin ligaments lead to the archetype BAM with living hinges. In summary, this work provides a generalized metamaterial architecture that allows expanding the palette of materials to select for BAM, from elastomers to elastic-perfectly plastic materials with further possibilities of using brittle materials.

\section{CONCLUSIONS}
This paper has proposed a generalization of the original architecture of bistable auxetic metamaterials (BAM) to allow their realization out of rigid solids. The revised design is cellular and features slender compliant ligaments bridging gaps between rotating and stationary units. The ligament geometry has been tailored to further smoothen stress peaks that localize in the ligaments. Maps presented here for Delrin$^\text{\textregistered}$ Acetal Resin are provided to engineer nominal architecture for durable BAM that can fulfill given engineering requirements. This work expands the selection of raw materials for BAM from solely elastomers to rigid materials, such as solids with elastic-perfectly plastic constitutive law and potentially brittle materials. By doing so, the potential use for BAM can be foreseen for solar panels and turbo engines, which require high mechanical performance. This study can be also applied in future to unit cells with other shapes introduced with the original BAM~\cite{Rafsanjani2016}.

\section*{ACKNOWLEDGEMENTS}
A.R. also acknowledges the financial support provided by Swiss National Science Foundation (SNSF) under grant number 164648.

% REFERENCES

\end{document}